\documentclass[%
 reprint,
 amsmath,amssymb,
 aps,
]{revtex4-1}
\usepackage{graphicx}
\usepackage{dcolumn}
\usepackage{bm}
\usepackage[colorlinks]{hyperref}
\usepackage[utf8x]{inputenc}
\usepackage{braket}
\usepackage{amssymb}
\usepackage{amsmath}
\usepackage{graphicx}
\usepackage[colorinlistoftodos]{todonotes}
\usepackage{lipsum}

\begin{document}

\preprint{APS/123-QED}

\title{Quantum optics of nonlinear systems in cascade}

\author{G.Buonaiuto$^{\dagger}$}
\email{gbuonaiuto1@sheffield.ac.uk}
\author{E.Cancellieri$^{*}$}
\author{D.M. Whittaker$^{\dagger}$}%
\affiliation{$^{\dagger}$Department of Physics and Astronomy, University of Sheffield, Sheffield S3 7RH, United Kingdom}
\affiliation{$^{*}$Department of Physics, University of Lancaster, Lancaster LA1 4YB, United Kingdom}%
\begin{abstract}
In this letter, we investigate the quantum optical properties of
driven-dissipative nonlinear systems in a cascade configuration. We show that pumping a nonlinear system with a state having a non-coherent statistics, can improve the antibunching of the output state and, consequently, the non-classicality of the whole system. Furthermore, we show that is possible to generate entanglement through dissipative coupling. These results applies to a broad category of physical systems with a Kerr-like non-linearity, from Rydberg atoms to exciton polaritons in microcavities.
\end{abstract}

\date{\today}

\maketitle

   

\textit{Introduction}. The ability of generating nonclassical states of light is a key requirement for performing quantum information processing using photons \cite{kok}. Squeezed states and EPR photon pairs, can be used to perform a variety of scientific tasks beyond quantum computing, such as performing precision measurements \cite{Anis}, like detecting gravitational waves \cite{Goda2}. In recent years, huge efforts, both theoretical and experimental, have been devoted to investigate methods to realize such quantum states of light. For example, using Kerr nonlinearities in coupled microcavities, to realize the unconventional photon blockade \cite{Flay,Sav}, or the generation of entangled states of light \cite{Liew1}. The use of others and more complex photonic structures \cite{Fara}, like quantum dots (ref), are examples of the range of possibilities contemplated in literature.

%

In this work we study the quantum optical properties of cascade nonlinear systems, where a nonlinear quantum system \textit{A} is coupled to another system \textit{B} through a Lindblad term. That is the second system is effectively driven by the first one. This kind of coupling was originally proposed by Gardiner \cite{Gard}, and adapted by Zoller \cite{Stann}, for two level systems in a waveguide. In particular, the possibility of performing universal quantum computing with coherent feedback networks of qubits has been recently demonstrated \cite{Pic}. Here we aim to provide the basics theoretical framework to implement \textit{cascade quantum computing} with continuous variable (CV) systems. The advantages of CV systems are that is much
easier to achieve higher detection efficiency, and the necessary states are 
generated in a deterministic way \cite{Thea}.

We aim to investigate the effect of the aforementioned effective driving on the output state, for two and three Kerr-nonlinear systems in cascade. In particular we focus on the change of photon statistics, non-Gaussianity and entanglement generation. Given the nature of the coupling, the system finds itself in a regime where a local breaking of the $U(1)$ symmetry is expected, which leads to a non stationary statistics, as have been shown in \cite{Buona}. First, we consider systems with strong Kerr nonlinearities, which are suitable to model the physics of optomechanical systems, like in \cite{Rabl}, where the appearance of nonclassical photon correlations in the combined strong coupling and sideband resolved regime has been demonstrated, and Rydberg atoms \cite{Peyr}, which can be used to realize strong nonlinear interactions between individual light quanta. Secondly, we consider the case of weakly non linear dissipative systems (like microcavity polaritons \cite{kav}), by setting the magnitude of the nonlinearity and of the effective dissipative rate to values comparable to the current experimental platforms based on III-V semiconductors.
In such setups, enhancing the non-classicality of the generated output can open a new route towards nonlinear quantum optics, which avoids single-atom strong coupling and trapping and where,instead, simple quasiparticles in semiconductor devices provide a source for quantum states of light.

\begin{figure}
\begin{center}
\includegraphics[width=\columnwidth]{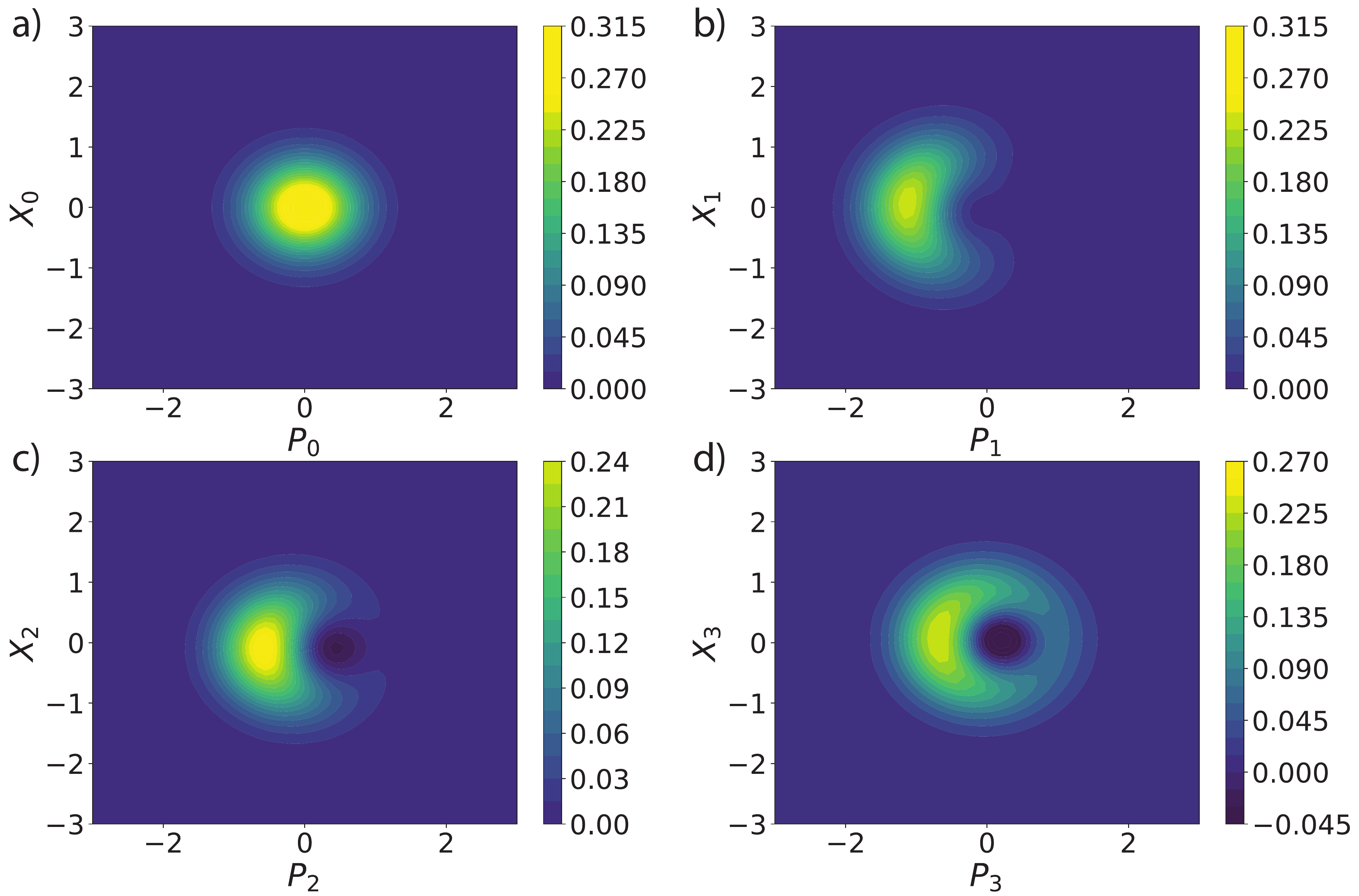}
   

  \label{fig:Ng2}
\caption{Wigner functions of the steady state for the coherent state \textbf{(a)}, representing the driving laser field, for the first \textbf{(b)}, the second \textbf{(c)} and the third \textbf{(d)} system in cascade, respectively, with $\delta=0$, $\gamma = 0.3$, $k=\gamma$ and pump intensity, $f=0.02\gamma$. The Wigner function becomes negative as we add more systems in cascade: the third mode \textbf{(c)} already shows a fidelity of 0.7 with a single photon state. }
\end{center}
\end{figure}

\textit{General model}.  An $N$-mode cascade  optical nonlinear system can be modeled following the approach of Gardiner et al. \cite{Gard}. The Hamiltonian of $N$ bosonic fields ($a_{m}$), with an Kerr nonlinearity $k_{m}$, one of them driven by a coherent pump with frequency $\omega_{P}$ and temporal shape $f(t)$, can be written as:
\begin{eqnarray}
H=\sum^{N}_{m=1}\omega_{m}a_{m}^{\dagger}a_{m}+ k_{m} a_{m}^{\dagger}a_{m}^{\dagger}a_{m} a_{m}+f(t) (e^{-i\omega_{P}t}a_{1}^{\dagger}+e^{i\omega_{P}t}a_{1}),\nonumber
\end{eqnarray}
where $\omega_{m}$ are the energies of each mode.
From the input-output theory it is possible to derive the master equation for the cascade system, where the field $a_{m}$ is driving the field $a_{j}$, with $j=m+1$:
\begin{eqnarray}
\label{master1}
\partial_{t}\rho =& &-i[H,\rho]+\sum^{N}_{m=1}\frac{\gamma_{m}}{2}(2 a_{m}^{\dagger}\rho a_{m}-a_{m}^{\dagger}a_{m} \rho+\rho a_{m}^{\dagger}a_{m})-\nonumber\\ & &\sum^{N}_{j=1}\sqrt{\eta_{mj} \gamma_{m}\gamma_{j}}([a_{j}^{\dagger},\rho a_{m}]+[\rho a_{m}^{\dagger},a_{j}]), 
\end{eqnarray}
where the first Lindblad term describes the dissipation for each mode, while the last one is the cascade driving of the $m$-th mode into the $j$-th. The parameters, $0 \le \eta_{ij} \le 1$, take into account a variety of phenomena like back-scattering and absorption, that change the effective coupling between the systems. In the following we are considering identical systems with the same energy, $\omega_{i}=\omega$, with identical decay rates, $\gamma_{m}=\gamma_{j}=\gamma$, identical nonlinearity $k_{m}=k_{j}=k$, and a perfect coupling between the modes in cascade ($\eta_{mj}=1$).

\textit{Wigner function}. To analize the effect of the cascade coupling, we evaluate numerically the time evolution of the Wigner function \cite{moya} for three systems in cascade, with a Quantum Monte Carlo approach \cite{joha}, expanding the total system on a properly truncated Fock basis. 
In Fig. 1 it is shown the result of the numerical analysis, when a steady-state is reached. Starting from the coherent state injection of the laser source, the first driven mode shows, as expected \cite{knight}, a crescent like shape, which is a signature of intensity squeezing. Moreover, the Wigner function for the second mode is also slightly negative, which indicates that is non-Gaussian. Finally, the state is moving from the intensity squeezing of the first system to a quasi-number state, as indicated by the increasing of the curvature of the probability distribution. To confirm this observation, it is possible to evaluate the fidelity $\mathcal{F}=Tr(\sqrt{\sqrt{\rho}\sigma \sqrt{\rho}})^{2}$ , between the cascade mode and a pure single photon state \cite{Niel}, where $\rho$ is the single-photon density matrix, $\sigma=\ket{1}\bra{1}$ and $\rho$ is the density operator of the cascade mode. The fidelity is one if the two systems are described by the same density matrix, while zero if they are completely orthogonal. While the first mode in cascade has a fidelity of $\mathcal{F}\approx 0.5$, for the second one we get $\mathcal{F}\approx 0.7$. This trend of increasing non-Gaussianity is confirmed by the Wigner function for the third mode in cascade: the distribution is more negative, compared to the previous case, and the fidelity between the mode and the single photon state is $\mathcal{F}\approx 0.8$.
   


   

\textit{Two cavities in cascade}. In order to get a better physical understanding of the cascade coupling on the statistics of the output field, we consider two Kerr modes:
\begin{eqnarray}
H=& &\omega a_{1}^{\dagger}a_{1}+\omega_{2}a_{2}^{\dagger}a_{2} + k a_{1}^{\dagger}a_{1}^{\dagger}a_{1} a_{1}+k a_{2}^{\dagger}a_{2}^{\dagger}a_{2} a_{2} +\nonumber \\ & &f(t) (e^{-i\omega_{p}t}a_{1}^{\dagger}+e^{i\omega_{p}t}a_{1}).
\end{eqnarray}
Following \eqref{master1}, the master equation for the two modes in cascade is:
\begin{eqnarray}
\label{master2}
& &\partial_{t}\rho =-i[H,\rho]+\frac{\gamma}{2}(2 a_{1}^{\dagger}\rho a_{1}-a_{1}^{\dagger}a_{1} \rho+\rho a_{1}^{\dagger}a_{1})+\nonumber \\ & &\frac{\gamma}{2}(2 a_{2}^{\dagger}\rho a_{2}-a_{2}^{\dagger}a_{2} \rho-\rho a_{2}^{\dagger}a_{2})- \gamma([a_{2}^{\dagger},\rho a_{1}]+[\rho a_{1}^{\dagger},a_{2}]).
\end{eqnarray}

To quantify the relevant statistical quantities associated to the cascade system, we make use of the positive-P representation \cite{carm}, which allows to convert the operator valued equation in a system of stochastic differential equations for c-numbers, $(a_{1}, a_{2})\Rightarrow (\alpha_{1}, \alpha_{1}^{*},\alpha_{2},\alpha_{2}^{*})$ (See Supplementary informations for more details). 
The first step in our analysis is to study the classical solution that can be obtained by taking the ensemble averages:  $\braket{\alpha_{1}}= S_{1}$ and $\braket{\alpha_{2}}=S_{2}$. Transforming the fields, $S_{1}\rightarrow S_{1}e^{-i \omega_{P}t}$ and $S_{2}\rightarrow S_{2}e^{-i \omega_{P}t}$ one can get rid of the time dependence in the pump term, and get the following mean-field equations:
\begin{equation}
\begin{cases} \partial_{t}S_{1}=-\gamma S_{1} -i \delta S_{1} -i f -i k |S_{1}|^{2}S_{1} \\ \partial_{t}S_{2}=-\gamma S_{2} -i \delta S_{2} -i k |S_{2}|^{2}S_{2}+\gamma S_{1}\end{cases}
\end{equation}
with $\delta=\omega-\omega_{P}$ being the laser-cavity detuning. Since we are interested in the steady state, the time derivative is set to zero and the classical equations simplifies to:(metti a sistema)
\begin{eqnarray}
&& \label{eq11}-i\gamma S_{1} + (\delta_{0} +k |S_{1}|^{2})S_{1}= - f \nonumber \\ &&\label{eq22}
-i \gamma  S_{2} + (\delta_{2} +k |S_{2}|^{2})S_{2}= -i \gamma S_{1}.
\end{eqnarray}
The first equation is completely independent from the second one, and it gives the typical bistable behavior \cite{Abr}, as a function of the pump intensity, provided that $\delta>\sqrt{3}\gamma$. The second equation is effectively driven by the first field, with an amplitude $i\gamma S_{1}$, so it can undergo a doubly bistable behavior: one inherited from the multiple solutions for $S_{1}$ and the other coming from  the intrinsic nonlinearity of the second system, as shown in Fig. 2. It can be noticed indeed that the second system has a bistable jump in the classical solution exactly when the first system itself is bistable, and another one, more prominent, which comes out from the intrinsic properties of the second system in cascade.
\begin{figure}
\begin{center}
   \includegraphics[scale=0.26]{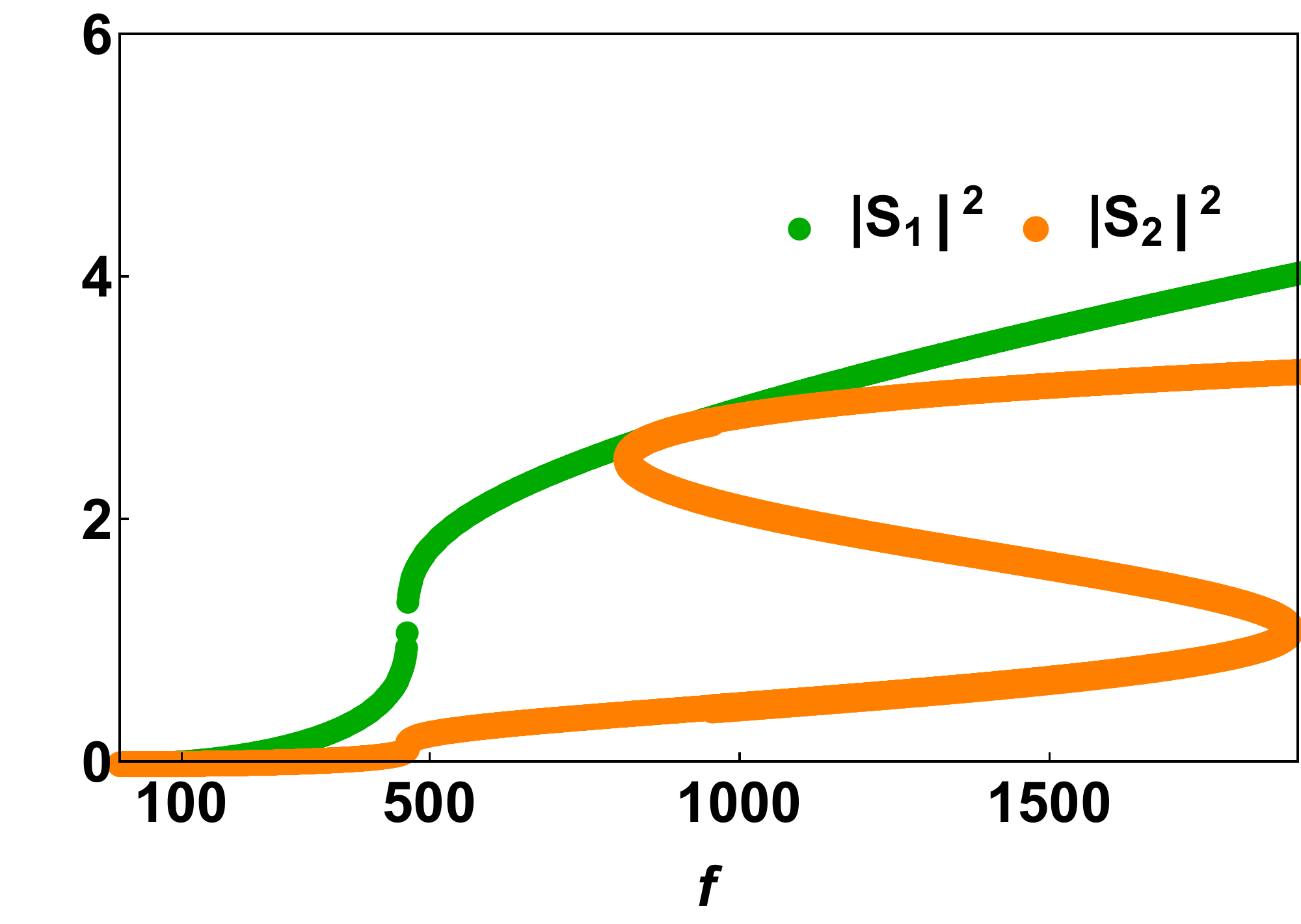}
   

  \label{fig:Ng2}
\caption{Bistable and multi-stable behaviour in the number of particles, as function of the pump intensity, with $\gamma=0.5$, $k=0.05$, $\delta=2*\sqrt{3}\gamma$.}
\end{center}
\end{figure}

To evaluate the quantum statistical quantities of interest it is necessary to consider the fluctuation around the classical solutions. For this study it is convenient to consider only the case for $\delta<\sqrt{3}\gamma$, where there is no bistable behavior. To this end, it is convenient to scale the fluctuations by the classical solutions, $\alpha_{1} \rightarrow S_{1}(1+\tilde{\alpha_{1}})$ and $\alpha_{2} \rightarrow S_{2}(1+\tilde{\alpha_{2}})$, and expand the fluctuations themselves in polar coordinates, $\alpha_{1} \approx S_{1}(1+i_{1}-i\theta_{1})$. Details about the linearisation and the fluctuation analysis are provided in the Supplementary information.

Knowing the expression for the fluctuations, it is possible to calculate the covariance matrix $\sigma$ for the system and then the relevant statistical quantities \cite{walls}, like the second order correlation function, $g^{(2)}_{m}(0)$, with $m=1,2$ \cite{Davi}. In polar coordinate these functions assume the rather simple expressions:
\begin{equation}
g^{(2)}_{m}(0)=\frac{\braket{\alpha_{m}^{*}(t)\alpha_{m}^{*}(t)\alpha_{m}(t) \alpha_{m}(t)}}{\braket{\alpha_{m}^{*}(t)\alpha_{m}(t)}^{2}}\approx 1+4 \braket{i_{m}^{2}},
\end{equation}
where $\braket{i_{1}^{2}}$ and $\braket{i_{2}^{2}}$ represents the variances of the intensity fluctuations. The information about the variances are contained in $\sigma$, which is evaluated analytically, giving for the first mode the following expression for $g^{(2)}_{1}(0)$:

\begin{eqnarray}
\label{g21}
g^{(2)}_{1}(0)=1-k\frac{\delta+k |S_{1}|^{2}}{\gamma^{2}+(\delta+k |S_{1}|^{2})(\delta+3k |S_{1}|^{2})}.
\end{eqnarray}
A general look to equation \eqref{g21}, reveals some details about the deviation from a coherent statistics due to the nonlinearity: if $\delta\le 0$, increasing the population pulls the system closer to resonance, so bunching is expected. On the contrary, when $\delta \ge 0$, increasing the population pushes the system away from resonance, leading to antibunching.

The expression for $\braket{i^{2}_{2}}$, needed to calculate $g^{(2)}_{2}(0)$, is long and not-straightforward: it is then more explicative, in order to visualize the effect of the cascade coupling on the correlation function, to analyze the results for a range of parameters.  
It is worth noting here that, from equation \eqref{eq11} one can express the population in the first mode, $|S_{1}|^{2}$, as a function of the population in the second mode:$|S_{1}|^{2}=(1+\frac{1}{\gamma^{2}}(\delta+k|S_{2}|^{2})^{2})|S_{2}|^{2}$. Therefore, as the $g^{(2)}_{m}$ depend only on the populations and not on the fields' amplitude, we can study them only as a function of $|S_{2}|^{2}$.
In Fig. 3 we show the correlation functions for each of the two cavities as functions of the populations in the second cavity and of the detuning from the pump, with a fixed set of parameters.
\begin{figure}
\begin{center}
   \includegraphics[width=\columnwidth]{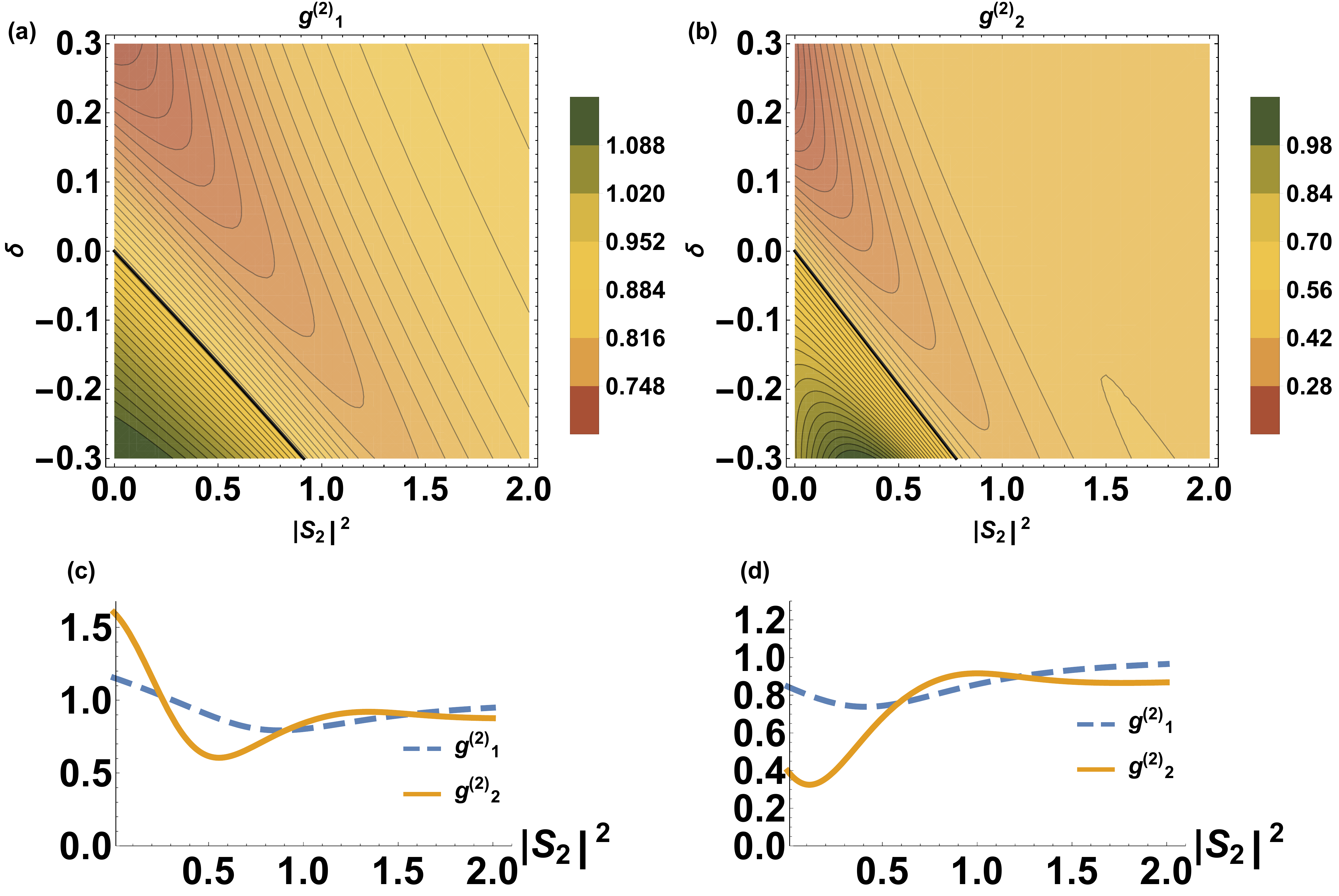}
   

  \label{fig:Ng2}
\caption{\textbf{(a)} $g^{(2)}_{1}$ and \textbf{(b)} $g^{(2)}_{2}$ for the cascade system as a function of the population in cavity two and of the detuning, with $\gamma=0.2$, $k=\gamma$. The oblique black curve corresponds to $g^{(2)}_{m}=1$, while the highlighted region are the ones where the system is antibunched. The correlation function for the first cavity has a minimum value of 0.65, while for the second cavity it goes down to 0.2.\textbf{(b-c)} Linear section of the contour plot, for $\delta=\pm 0.12$, respectively.}
\end{center}
\end{figure}
For $\delta >0$, both modes show sub-Poissonian statistics and for small values of $|S_{2}|^{2}$ one can identify a region of the phase-space where the second mode shows a strongly improved antibunching, with respect to the first one. In particular, considering the situation shown in Fig. 3, it is possible to notice that the minimum value of the second order correlation function for the first mode is $min(g^{(2)}_{1})=0.65$, while for the second one is $min(g^{(2)}_{2})=0.2$. 

\textit{Three cavities in cascade}. It is interesting to study the possibility whether the gain of the antibunching follows an extensive principle: i.e. if having more systems in the cascade results in an improved value of the quantum correlations.\begin{figure}
\begin{center}
   \includegraphics[width=\columnwidth]{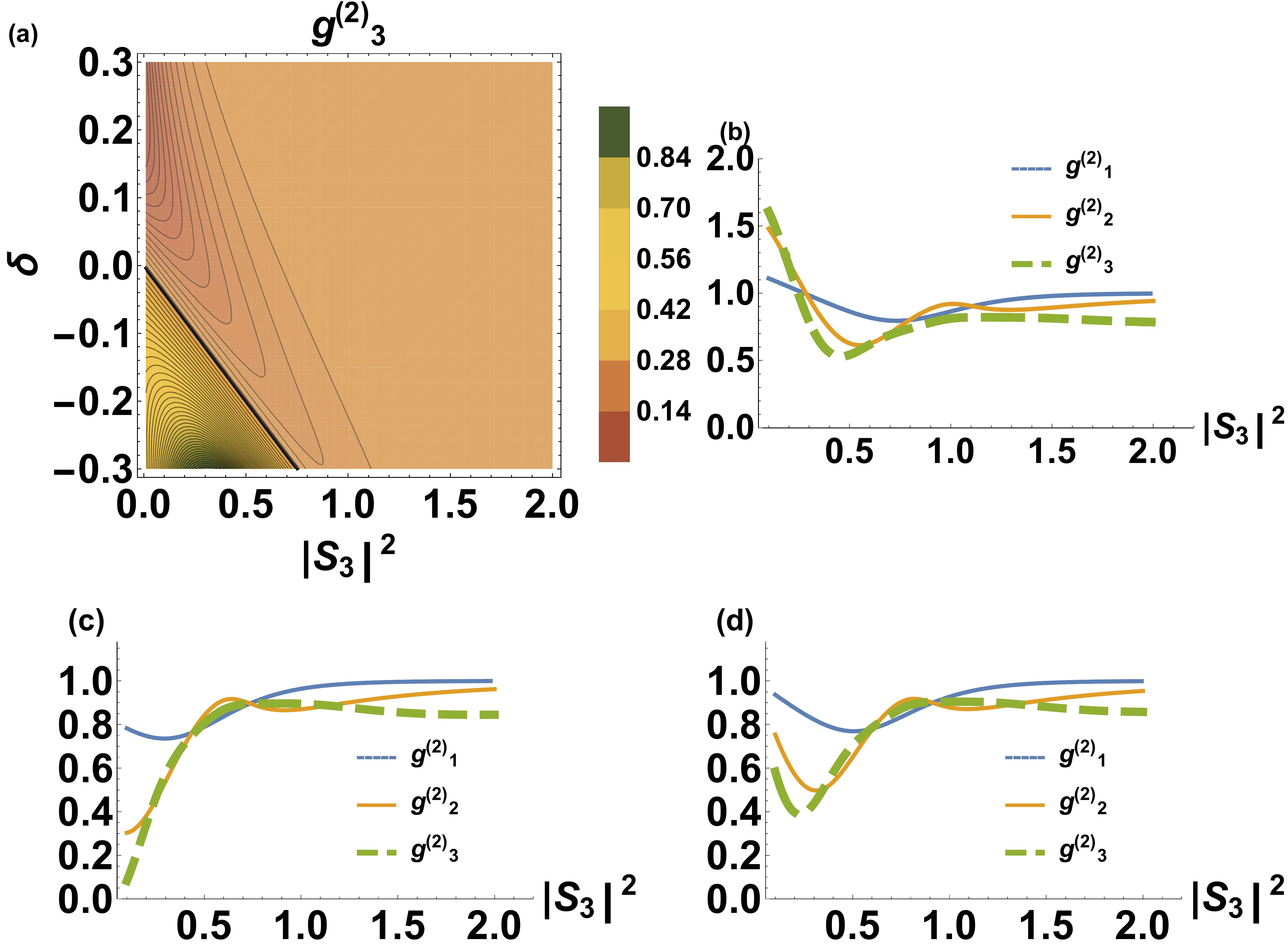}
   

  \label{fig:Ng2}
\caption{\textbf{(a)} $g^{(2)}_{3}$ for the cascade system as a function of the population in mode 3 and of the detuning, with $k=\gamma=0.2$. The black curve corresponds to the values for which $g^{(2)}_{3}=1$, while the highlighted region are the ones where the system is antibunched. The third mode in cascade shows a further improvement of the antibunching, in particular, for this set of parameters, $min(g^{(2)}_{3}=0.05$. \textbf{(b)} Section of the phase space for the correlation functions for the three modes, with $\delta=-0.12$, \textbf{(c)} $\delta=0$, \textbf{(d)}  $\delta=0.12$.}
\end{center}
\end{figure}
However, given the nonlinear nature of the equation we are considering it is not an easy task to demonstrate this analytically, therefore we limit the our study to the case of three nonlinear systems in cascade using the same semi-analytical procedure adopted in the previous section. The second order correlation function for the third mode is shown in Fig. 4. As before, it is possible to perform this analysis as a function of the population in the last mode in cascade, since $|S_{2}|^{2}=(1+\frac{1}{\gamma^{2}}(\delta+k|S_{3}|^{2})^{2})|S_{3}|^{2}$ (See Supplementary materials for details).
As shown in Fig. 4, the minimum value of the correlation function for the third mode, $min(g^{(2)}_{3}) = 0.05$, which is a further improvement, compared to the two mode configuration.

\textit{Two modes entanglement through dissipation}. In analogy to Bell’s result for discrete variable entanglement, the presence of CV entanglement
between mode one and two, can be characterized studying \cite{Duan}:
\begin{equation}
\label{ent}
E_{N}=V(p_{1}-p_{2})+V(x_{1}+x_{2}),
\end{equation}
where $x_{1}=(a_{1}+a^{\dagger}_{1})/2$ and $p_{1}=(a_{1}-a^{\dagger}_{1})/2i$, and $V$ is the variance. If $min_{\phi}(E_{N})=\tilde{E}_{N}<1$, where $\phi$ is the relative phase between the two modes, the system is in an entangled state, while for $\tilde{E}_{N}\geq 1$ they are separable. 
In this case, the minimum value of $E_{N}$ is found for $\phi=\pi/2$, and $\tilde{E}_{N}$ can be written as:
\begin{eqnarray}
&&\tilde{E}_{N}=1+\braket{a^{\dagger}_{1}a_{1}}+\braket{a^{\dagger}_{2}a_{2}}-\braket{a^{\dagger}_{1}}\braket{a_{1}}-\braket{a^{\dagger}_{2}}\braket{a_{2}}-\nonumber\\ &&2\sqrt{\braket{a^{\dagger}_{1}a^{\dagger}_{2}}-\braket{a^{\dagger}_{1}}\braket{a^{\dagger}_{2}}}\sqrt{\braket{a_{1}a_{2}}-\braket{a_{1}}\braket{a_{2}}}.
\end{eqnarray}

The expectation values entering this equation can be calculated following the same approach of the previous sections, i.e. using the mean field equation and the covariance matrix to evaluate the correlations. Explicitly, in terms of the polar expansion, the bipartite CV entanglement witness reads:
\begin{eqnarray}
&&E^{i,\theta}_{N}=1+|S_{1}|^{2}\braket{i^{2}_{1}}+|S_{2}|^{2}\braket{i^{2}_{2}}-2|S_{1}|\times|S_{2}|\nonumber\\ &&\sqrt{(\braket{i_{1}i_{2}}-\braket{\theta_{1}\theta_{2}})^{2}+(\braket{i_{1}\theta_{2}}+\braket{i_{2}\theta_{1}})^{2}}.
\end{eqnarray}
As shown in Fig. 5(a), for low values of the population (i.e. dark region on the left side) the two modes are in an entangled state. This is particularly interesting, because it shows that entanglement can be generated through a dissipative coupling. This is potentially relevant for quantum communication protocols, in particular for generating entanglement between systems which are spatially separated \cite{krau}, and for quantum foundation experiments, like measuring Bell's inequalities \cite{Thea} with CV systems. From Fig. 5, it can be noticed that the violation of the inequality quickly saturates, as the number of excitation in mode $2$ increases.
\begin{figure}
\begin{center}
   \includegraphics[width=\columnwidth]{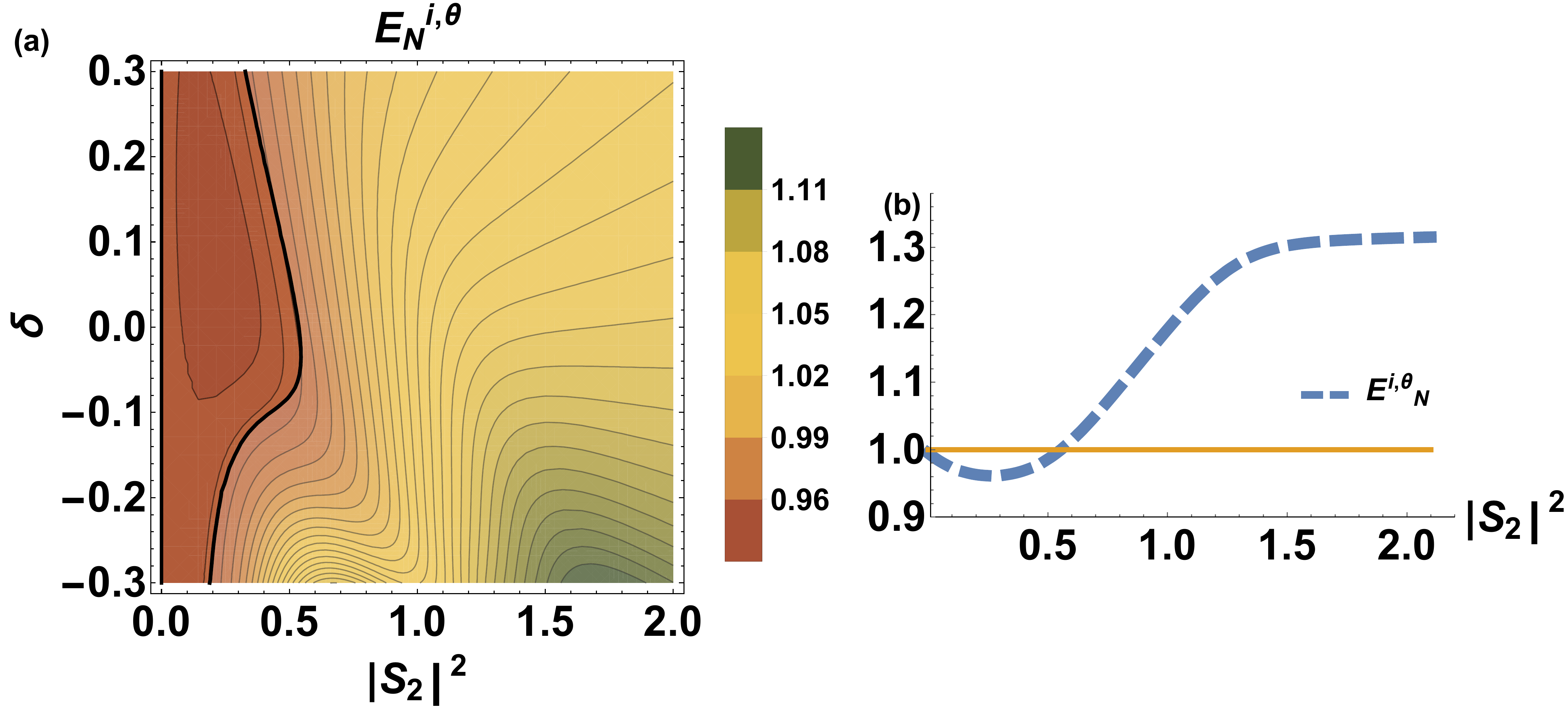}
   

  \label{fig:Ng2}
\caption{\textbf{(a)} Bipartite Entanglement between the modes, as a function of the population $|S_{2}|^{2}$ and of the detuning $\delta$, with $\gamma=0.2$, $k=\gamma$. The black curve specifies the point where the inequality saturates to 1 while the highlighted region corresponds to values where the systems are effectively entangled.\textbf{(b)} A 1D section of the 2D plot which underlines the range of the violation of the inequality, with $\delta=0$.}
\end{center}
\end{figure}
Indeed, at that point, the occupation number for the two modes start to be significantly different, which means that in the number operator basis the two systems become increasingly distinguishable (separable). It is worth to notice here that the maximum degree of entanglement does not change as a function of the onsite Kerr nonlinearity (as shown in the Supplementary material), but is rather a function of the effective coupling strength between the cascade modes. The nonlinearity simply shifts the driving field value for which the minimum of the entanglement witness occurs.

\textit{Weak nonlinear system}. So far we have considered systems with a strong on-site Kerr nonlinearity, $k = \gamma $. However, this is not the case in many realistic experimental setup, like III-V semiconductor microcavity-polariton systems.
\begin{figure}
\begin{center}
   \includegraphics[width=\columnwidth]{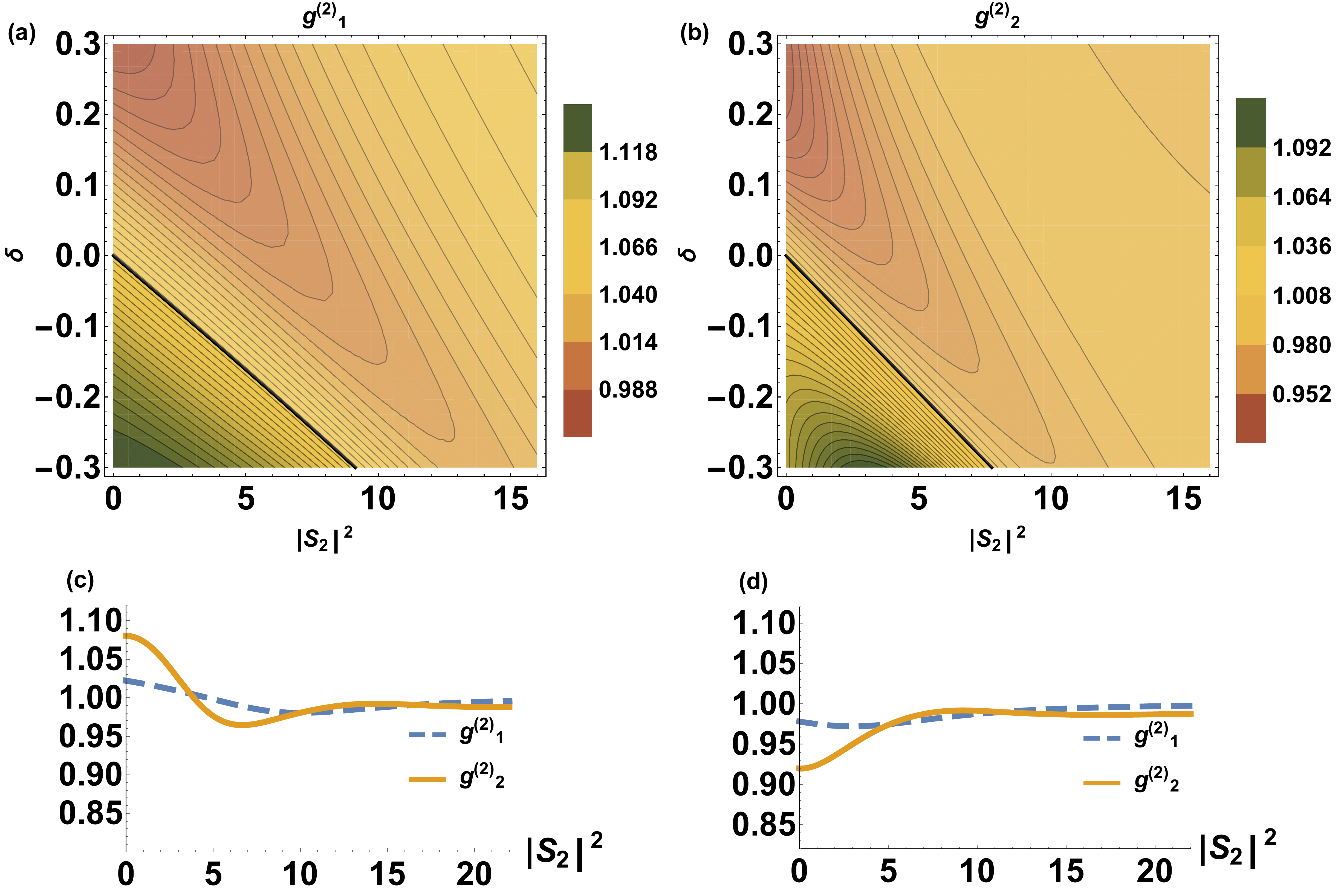}
   

  \label{fig:Ng2}
\caption{\textbf{(a)} $g^{(2)}_{1}$ and \textbf{(b)} $g^{(2)}_{2}$ for the cascade system as a function of the population in cavity two and of the detuning,with $\gamma=0.2$, in the weak nonlinearity regime, $k=0.005$. The correlation function for the first cavity is bounded from below to 0.98, the second cavity shows an improvement of the sub-Poissonian statistics, down to 0.95.}
\end{center}
\end{figure}
In such system the strength of the nonlinearity is order of magnitude smaller than the linewidth, $k\approx \gamma 10^{-2}$. Nevertheless, is still interesting to investigate quantum effects in these systems \cite{Lop}. In this systems, since the nonlinearity is small, the deviation from a coherent statistics is almost negligible and therefore, is interesting to study the cascade configuration, to enhance the quantum effects. For example, as shown in Fig. 6 (a), a single Kerr quantum system with $k=0.005$ and $\gamma=0.2$, reveals an optimal value for the $g^{(2)}_{1}(0)\approx 0.98$, as expected from Drummond et al. \cite{Drumm}. Using a further system in cascade, is possible to achieve a lower value for the antibunching, down to $g^{(2)}_{2}(0)\approx 0.95$, as shown in Fig.6 (b).

\textit{Conclusion}. In this work we have investigated the quantum optical properties of nonlinear cascade systems, showing that the effective driving generated by an incoherent coupling between Kerr oscillators, is effecting the correlation functions and the non-classicality of the output state. Moreover, we also showed that a dissipative coupling, can give rise to entangled states, which can be used to implement quantum computing protocols and quantum foundations tests with CV systems. Finally, we showed that such effect holds also for weakly non-linear systems, paving the way to investigate experimentally quantum phenomena in mesoscopic semiconductor-based technologies.
\\
\bibliographystyle{ieeetr}
\bibliography{pap2}
\end{document}